\documentclass[11pt]{article}

\usepackage{graphicx}
\usepackage[dvips]{color}
\usepackage{mathtext}
\usepackage{cite}
\newcommand{\kmng}{K_{\mu 2\gamma}}
\newcommand{\kmnp}{K_{\mu 3}}

\date{}

\begin{document}

\title{First measurement of the T-violating muon polarization in 
the decay $K^+\rightarrow\mu^+\nu\gamma$ }

\author{V.V. Anisimovsky$^a$\thanks{Corresponding author.
{\it Email address:} valera@al20.inr.troitsk.ru (V.V. Anisimovsky).}, 
A.N. Khotjantsev$^a$, A.P. Ivashkin$^a$, M. Abe$^b$, \\
M. Aliev$^a$, M. Aoki$^c$, Y. Asano$^b$, M. Blecher$^d$, \\
P.~Depommier$^e$, M. Hasinoff$^f$, K. Horie$^g$, H.C. Huang$^h$, \\
Y. Igarashi$^g$, J. Imazato$^g$, M.M.~Khabibullin$^a$,  \\
Yu.G. Kudenko$^a$, Y. Kuno$^c$, A.S. Levchenko$^a$, G.Y. Lim$^g$,  \\
J.A. Macdonald$^i$, O.V. Mineev$^a$, N.~Okorokova$^a$, \\
C.~Rangacharyulu$^j$, S.~Shimizu$^c$, Y.-M.~Shin$^i$, N.V.~Yershov$^a$, \\
T. Yokoi$^g$ \\
 ~\\ 
(KEK E246 Collaboration)
 ~\\
 ~\\
$^a${\it Institute for Nuclear Research RAS, 117312 Moscow, Russia} \\  
$^b${\it Institute of Applied Physics, University of Tsukuba, Tsukuba, 305-0006, Japan} \\  
$^c${\it Osaka University, Osaka, 560-0043, Japan} \\
$^d${\it Virginia Polytechnic Institute and State University, VA 24061-0435, USA} \\
$^e${\it Universit\'e de Montr\'eal, Montr\'eal, H3C 3J7 Canada} \\
$^f${\it University of British Columbia, Vancouver,  V6T 2A3 Canada} \\
$^g${\it High Energy Accelerator Research Organization (KEK), Tsukuba, 305-0801 Japan} \\
$^h${\it Department of Physics, National Taiwan University, Taipei 106, Taiwan} \\
$^i${\it TRIUMF, Vancouver, V6T 2A3 Canada} \\
$^j${\it University of Saskatchewan, Saskatoon,  S7N 5E2 Canada} \\
}

\maketitle

\begin{abstract}
We present the result of the first measurement of the 
T-violating transverse muon polarization $P_T$
in the decay $K^{+}\rightarrow\mu^{+}\nu\gamma$.  This polarization 
is sensitive to new sources of CP--violation in the Higgs sector. 
Using data accumulated in the period 1996-98 we have obtained 
$P_T=(-0.64 \pm 1.85(stat) \pm 0.10(syst))\times 10^{-2}$  which is 
consistent 
with no T-violation in this decay. 

\end{abstract}

\section{Introduction}

The decay $K^+\to\mu^+\nu\gamma$ ($\kmng$) is one of the kaon decays 
which 
could  probe  new physics beyond the Standard Model (SM). However, 
the decay is relatively poorly measured and there have been only a few  
experiments 
that determined the $\kmng$  branching 
ratio~\cite{exp_kek85,exp_itep89} 
and the structure-dependent component~\cite{exp_bnl00}. Other 
interesting observables for this decay are  the components of the muon 
polarization:  $P_T$ which 
is the T-odd  polarization component normal to the decay plane defined 
as $P_T=\vec{s}_\mu\cdot
(\vec{p}_\gamma\times\vec{p}_\mu)/|\vec{p}_\gamma\times\vec{p}_\mu|$, 
and $P_N$--normal muon polarization
(T--even) in the decay plane which is sensitive to  the values and 
signs of the 
kaon vector  $F_V$ and axial-vector $F_A$ form factors, 
as shown in~\cite{bgk}.
The contribution to $P_T$ from the Standard Model is known to 
be very small $\simeq 10^{-7}$~\cite{bigi}.
The T-even transverse polarization due to final state
interactions (FSI) can mimic the T-odd effect at the $\leq 10^{-3}$ 
level~\cite{Efrosinin:yt,fsi_braguta,fsi_rogalyov}. 
However, it can be reliably calculated. Therefore, 
any  non-zero value of $P_T$ above the level of FSI  or  
different from  the FSI values, if the experimental  sensitivity to 
the polarization is   better than $10^{-3}$,    would indicate
new physics. Several non-SM 
models predict non-zero values for $P_T$ in
$\kmng$ 
decay~\cite{pt_kobayashi,pt_chen,pt_wu} 
and from two of them (SUSY with R-parity and multi-Higgs models 
without natural flavor 
conservation) large values of $P_T\leq 10^{-2}$ are expected.

In this letter we present the first results  on  $P_T$ obtained 
in the analysis of the $\kmng$ data accumulated in the 
KEK E246 experiment for the
1996--98 data taking period.

\section{Experiment}

\begin{figure}[htb]
\centering\includegraphics[width=14cm]{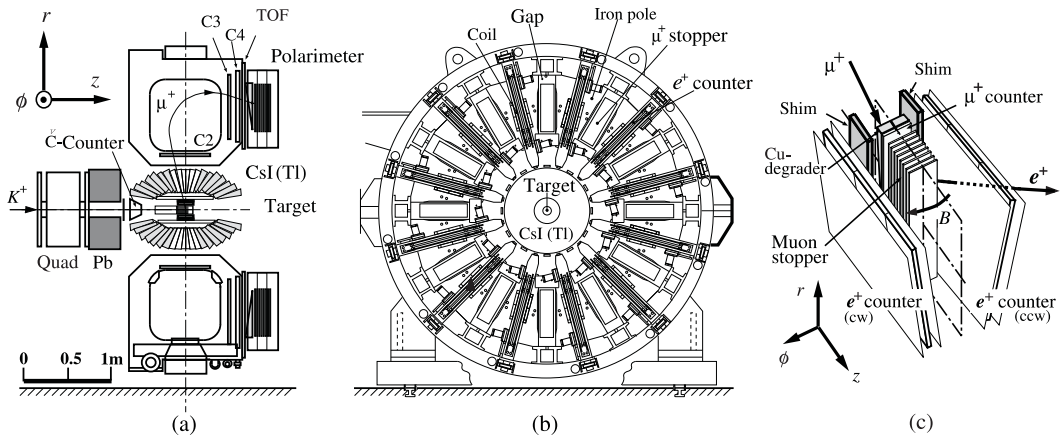} 
\caption{The layout of the KEK E246 detector:(a) side view, (b) end 
view and (c) one sector of the polarimeter.}
\label{fig:e246setup}
\end{figure}

The measurement was performed using the detector constructed to search 
for T-violating muon polarization in the $K^+\to\pi^0\mu^+\nu$ 
($\kmnp$) decay. The apparatus is shown in Fig.~\ref{fig:e246setup}, 
and described in detail 
elsewhere~\cite{cherenkov,hodoscope,e246setup,csi}.
A separated $K^{+}$ beam ($\pi/K \simeq 6$) of 660 MeV/$c$ was used 
with a typical intensity 
of $3.0\times 10^5$ kaons per 0.6-s spill duration with a repetition 
of 3 s.
A ${\rm \check{C}}$erenkov counter   with a multiplicity trigger 
distinguished $K^{+}$s
 from $\pi^+$s
 with an efficiency of more than 99\%. Kaons 
then were   
slowed in a Al+BeO degrader, and stopped in an active  target made 
of 256 
scintillating fibers 
located at 
the center of a 12-sector superconducting toroidal spectrometer. A 
charged 
particle ($\mu^+$, 
$e^+$ or $\pi^+$) from the kaon decay 
hit one of the twelve fiducial counters 
surrounding the target
and was
bent through
one of the  spectrometer sector gaps
 and
detected by the
 TOF and $\mu^+$ 
counters at the entrance of the polarimeter.
   Tracking of charged particles was provided by the
target fibers, a scintillation ring hodoscope   surrounding the target 
and 
multiwire 
proportional chambers at the entrance (C2) and exit (C3 and C4) of 
each magnet sector gap.
The momentum resolution of the whole tracking system was 
$\sigma_p$=2.6 MeV/$c$ at 
205~MeV/$c$. 

The energies and angles of the photons  were measured 
by  the  CsI photon 
calorimeter consisting of 768  CsI(Tl) modules \cite{csi}. The 
CsI(Tl) barrel has 
twelve holes 
for the charged particle  entry to  the toroidal spectrometer and 
covers a solid 
angle of about $0.75\times 4\pi$~sr.  Energy resolution of 
$\sigma_E/E=2.7$\% at 
200 MeV and 
angular resolutions of $\sigma_{\theta} = 2.3^{\circ}$ were obtained.
Good time resolution of 3.5~ns~(rms) at 100 MeV allowed us to use 
the fast CsI 
signal at the
trigger level  and thereby effectively suppress accidentals in the 
calorimeter.

Muons entering the polarimeter (Fig.~\ref{fig:e246setup}c) were 
degraded by an 
Al+Cu block and stopped in a stack of pure Al plates. The stopper 
material (Al) was 
chosen because there is no depolarization of muons stopped in aluminum.
Positrons from  $\mu^+\to e^+\nu\bar{\nu}$  decays of stopped muons 
were 
detected by  12 positron counters
consisting of three layers of plastic scintillators 
which were located azimuthally
between the
muon stoppers.
The time spectra of the positrons were recorded by multi-stop TDCs.

A three-level trigger for selection of events was used. At the first 
level,
 a signal from the kaon ring in the ${\rm \check{C}}$erenkov detector 
followed by a hit in at least one  fiducial counter in coincidence 
with 
 signals from the
TOF and  $\mu^+$ counters of a corresponding magnet sector were 
 required. 
The second level trigger required at least one hit in the
 CsI(Tl) calorimeter, and 
the third level
 trigger required the triple coincidence signal from the 
positron counter arrays adjacent to the corresponding magnet 
sector within 
20~$\mu$s
after the first level trigger.

The T-violating asymmetry was extracted using a double ratio as:
\begin{equation}
A_T =\frac{1}{4} \left[ \frac{(N_{cw} / N_{ccw})_{fwd}} {(N_{cw} / 
N_{ccw})_{bwd}} - 1 \right] \label{atdr}
\end{equation}
Here, $N_{cw}$ and $N_{ccw}$ are the sums over all 12 sectors  
of counts of 
clockwise ($cw$) and 
counter-clockwise 
($ccw$) emitted positrons. Indices $fwd$ and $bwd$ denote 
two classes 
of events: forward  events ($fwd$) when the angle between the 
photon and the beam 
direction (z-axis) was 
less than $70^{\circ}$ and backward  events ($bwd$) when the 
angle between 
the photon and the beam direction
was more than $110^{\circ}$.
The signal values $N_{cw}$ and $N_{ccw}$ were extracted by 
integrating the 
positron time spectrum 
 in the  polarimeter 
after subtraction 
of the background. The signal integration region was chosen to 
be from 
20 ns to 6 $\mu$s and the background 
fitting  region for the background extraction was from 
6.0  to 19.5 $\mu$s.  

The sign of $A_{T}$ for events with
forward-going photons is opposite to that for events with 
backward-going 
photons. This allows us to employ a  double ratio method which 
is a powerful 
technique for reduction of most systematic errors.   Moreover, 
considerable 
reduction of systematic effects
 was achieved from 
the azimuthal symmetry of the 12-gap detector and the use 
of the same 
positron detector as 
a $cw$ for one gap and  a $ccw$ detector for the adjacent gap.

The value of $P_{T}$ is related to $A_T$ by
\begin{equation}
P_T = \frac{A_T}{\alpha \times f \times (1-\beta)}  \label{ptat}
\end{equation}
where $\alpha$ is the analyzing power of the polarimeter, 
$f$ is an angular 
attenuation factor and $\beta$ is the overall fraction of all 
backgrounds.

\section{Analysis}

The main problem 
of the $\kmng$ analysis is the rejection of the predominant $\kmnp$ 
background with only
one photon ($1\gamma$) from the $\pi^0$ decay  detected  by the CsI 
calorimeter. The extraction of $\kmng$ events was performed in two 
stages.

The first stage was mostly identical 
to that of the 
main ($\kmnp$) E246 analysis  and was used to suppress $K_{\mu2}$, 
$K_{\pi2}$, $K_{e3}$, 
 and  $K^+\to\pi^+\pi^+\pi^-$, $K^+\to\pi^+\pi^0\pi^0$ ($K_{\pi3}$)
background decays. 

\begin{figure}[htb]
\centering\includegraphics[width=12cm]{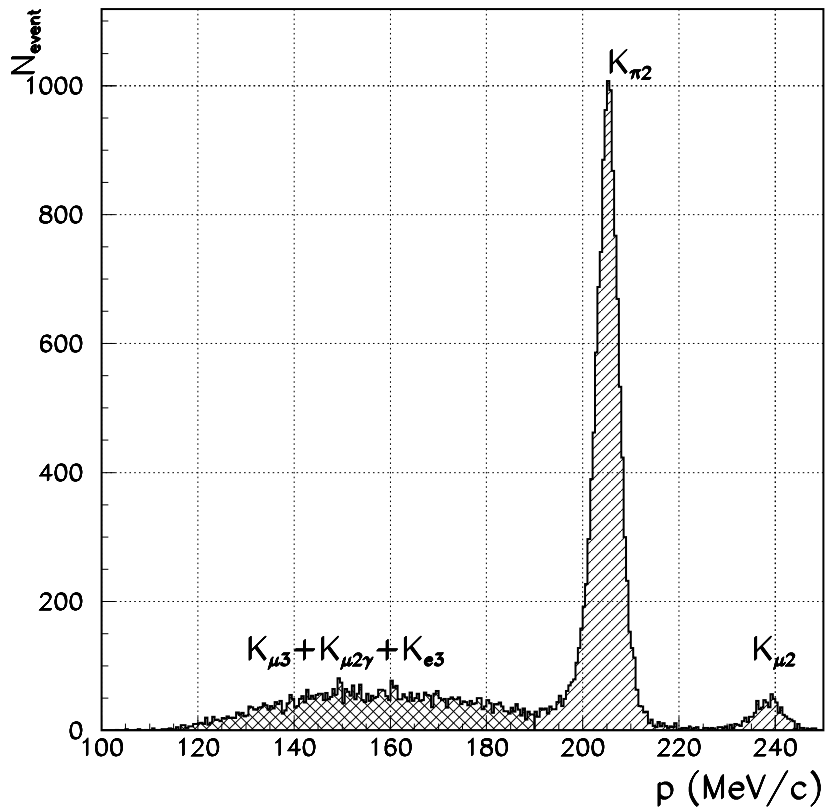} 
\caption{Momentum spectrum of charged particles for events with 
one photon detected by the CsI calorimeter. The double-hatched 
region shows the events with momentum less than 190 MeV/$c$.}
\label{e246ppi}
\end{figure}

\begin{figure}[htb]
\centering\includegraphics[width=12cm]{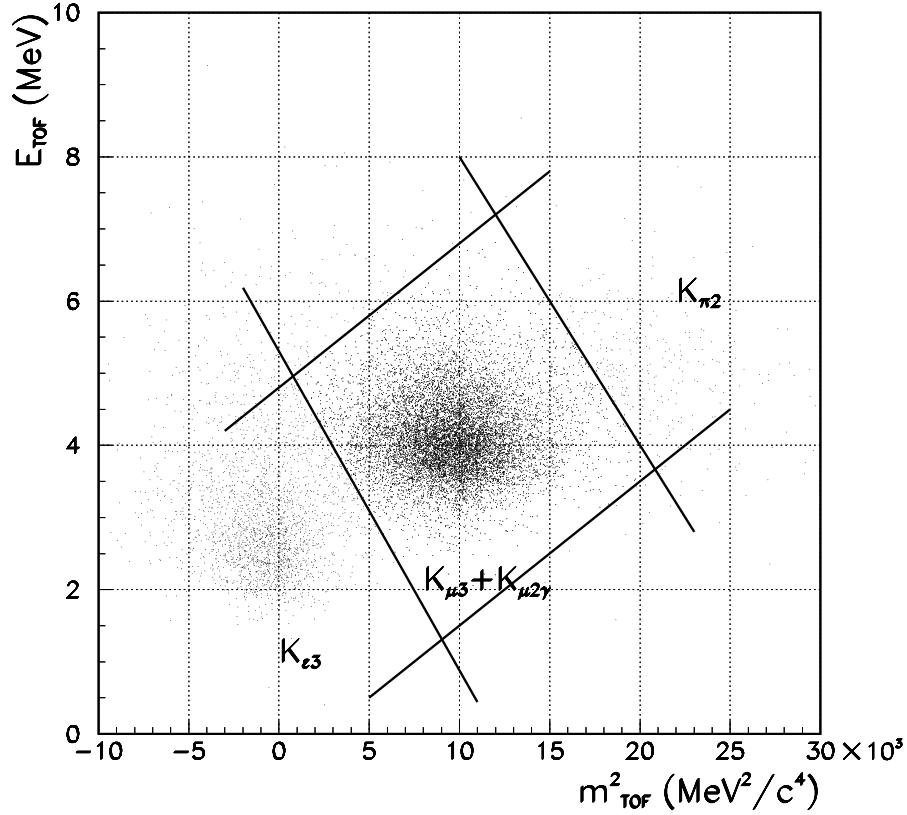} 
\caption{Rejection of $K_{e3}$ using the time-of-flight technique. 
The ``cloud'' in the 
bottom-left corner corresponds to positrons, the events inside the 
rectangle are muons.}
\label{e246m2tof}
\end{figure}

The $K_{\pi2}$ and $K_{\mu2}$ decays were rejected using a cut on 
the muon 
momentum   $ p_{\mu}\leq 190$ MeV/$c$ selecting the  
$K_{\mu2\gamma }$ muons
in  the region of 100-190 MeV/$c$, as illustrated in 
Fig.~\ref{e246ppi}.
Most of the muons from   $K_{\pi2}$ events, in which 
$\pi^+$ decays in-flight,  were rejected by 
 applying the  tracking cut $\chi^2<10$.  These events had kinks, 
i.e. the
tracks did not point back to the kaon vertex. 
The $K_{\pi3}$ backgrounds 
were negligible 
due to the combination of several suppressing factors: acceptance 
of the magnet 
spectrometer (the 
endpoints of charged particle momentum spectra for both decays 
are quite low: 
125 and 133 MeV/c, 
respectively), the low probability of detecting these decays as 
one-photon events and, finally, most 
of $\pi^+$s
 were stopped in the 
Cu degrader and 
did not reach the polarimeter. 
The $K_{e3}$ events were removed using a time-of-flight 
method by imposing a geometric cut on the
two-dimensional spectrum of a charged particle mass 
calculated from 
time-of-flight vs. the  energy deposited in the TOF 
counter (Fig.~\ref{e246m2tof}). 
The 
background from accidental photons in the CsI(Tl) detector 
was suppressed using 
a photon 
energy threshold of 50 MeV and by requiring a coincidence between
a photon 
signal in the CsI and a signal in the fiducial 
and TOF counters from a charged particle within a window of 
$\pm 15$~ns.
Background associated with the beam was suppressed by a veto 
counter system 
surrounding the beam region. 

\begin{figure}[htb]
\centering\includegraphics[width=12cm]{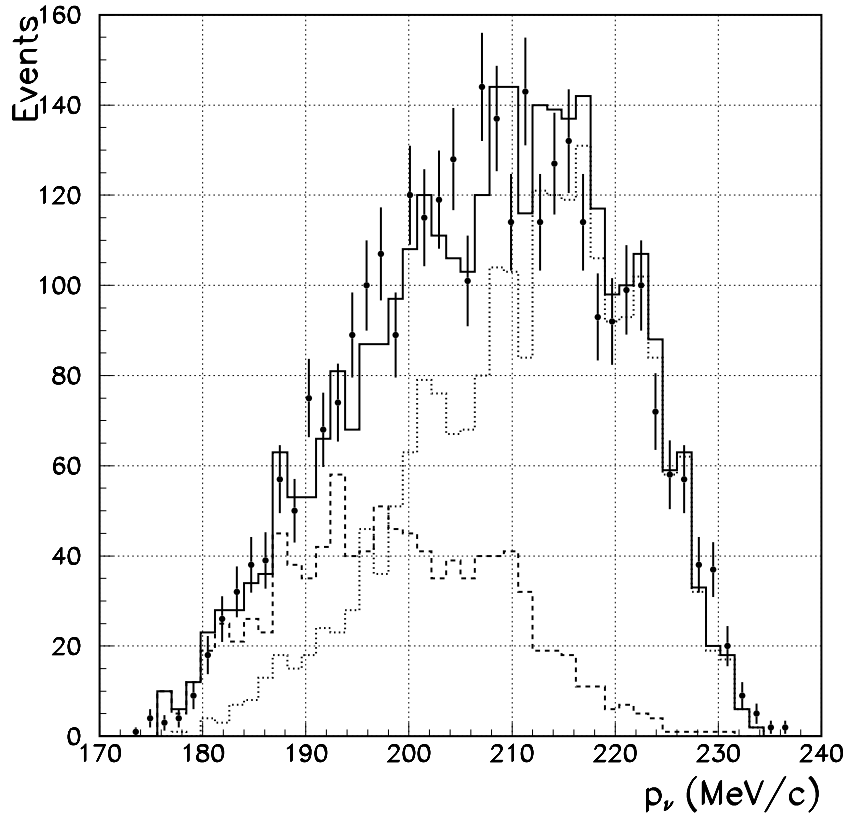} 
\caption{Momentum spectra of missing neutrino.
The black dots with error bars show the experimental 
data. MC simulation:
dotted  line~--~$K_{\mu2\gamma}$, dashed line~--~$K_{\mu3}$, 
solid line~--~$K_{\mu2\gamma} + K_{\mu3}$.
  Cuts:
$-0.7\times 10^4~{\mathrm MeV^2/c^4}<M_{miss}^2<1.5\times 
10^4~{\mathrm MeV^2/c^4}$,
$\Theta_{\mu\gamma}<90^\circ$.}
\label{kmngpnu}
\end{figure}

\begin{figure}[htb]
\centering\includegraphics[width=12cm]{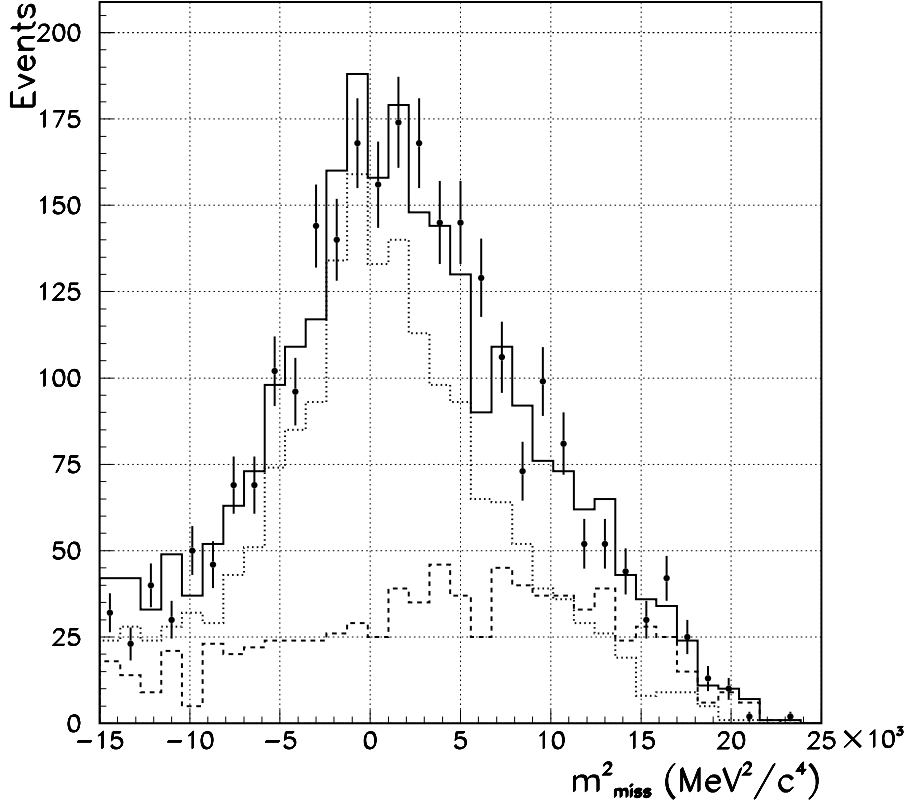} 
\caption{Missing   mass spectra for the following cuts:
$p_{\nu}>200~{\rm MeV/c}$,
$\Theta_{\mu\gamma}<90^\circ$. The black dots with error 
bars show the experimental data. MC simulation:
dotted  line~--~$K_{\mu2\gamma}$, dashed line~--~$K_{\mu3}$, 
solid line~--~$K_{\mu2\gamma} + K_{\mu3}$.}
\label{kmngmm2}
\end{figure}

The second stage of the analysis was aimed at the extraction of 
$\kmng$ events and 
the rejection of $\kmnp$ decays.  
Due to the   
 precise measurement of the muon momentum and the photon 
energy and direction, the kinematics  
of  $K_{\mu3}$ and  
$K_{\mu2\gamma}$ decays can be reconstructed 
completely~\cite{Kudenko:2000sc}.   The 
kinematic
 parameters such as missing mass squared
$M^2_{miss}$, the angle between the muon and the photon,   
$\Theta_{\mu\gamma}$,   
 and
the neutrino momentum, $p_{\nu}$,   
 can be efficiently used  
to suppress   
the $K_{\mu3}$ background. The missing mass squared 
for 1$\gamma$ events   
is
\begin{equation}
\label{missmass}
M^2_{miss} = E^2_{miss} - {\vec{p}}\,^2_{miss} = (m_K - E_{\mu} 
-E_{\gamma})^2 -
(\vec{p}_K - \vec{p}_{\mu} - \vec{q})^2,
\end{equation}
where $\vec{q}$ is the photon momentum.
For  $K_{\mu2\gamma}$ events, the neutrino is the only missing 
particle, therefore, $\vec{p}_{\nu} = \vec{p}_{miss}$, $E_{\nu} = 
E_{miss}$ and 
$M^2_{miss} = E_{\nu}^2 - {\vec{p}}\,^2_{\nu} = 0$, while the 
$M^2_{miss}$ of  
 $K_{\mu3}$ events is   
 distributed over a   
  wide range 
with the maximum  at
about $2\times 10^4$~MeV$^2$/$c^4$.  
 The reconstructed neutrino momentum of $1\gamma$ events from 
experimental 
and Monte Carlo (MC)
data after 
imposing the cuts 
$-0.7\times 10^4~{\rm MeV}^2/c^4 < M^2_{miss} < 
1.5\times 10^4~{\rm MeV}^2/c^4$ and 
 $\Theta_{\mu\gamma}\leq 90^{\circ}$, 
  and 
 accepting events with $E_{\gamma} > 50$ MeV is shown in 
Fig.~\ref{kmngpnu}.  
The peak at $p_{\nu}\sim 220$ MeV/$c$ in MC data   
corresponds to $K_{\mu2\gamma}$ 
events.   In the region with $p_{\nu} > 200$ MeV/$c$,  
 the contamination  of the $K_{\mu3}$ events is estimated to be 
$\simeq17$\%. 
Fig.~\ref{kmngmm2}
shows the  experimental and MC  spectra of  neutrino missing
mass after applying the cut ${p}_{\nu} > 200~{\rm MeV}/c$.  

\begin{figure}[htb]
\centering\includegraphics[width=12cm]{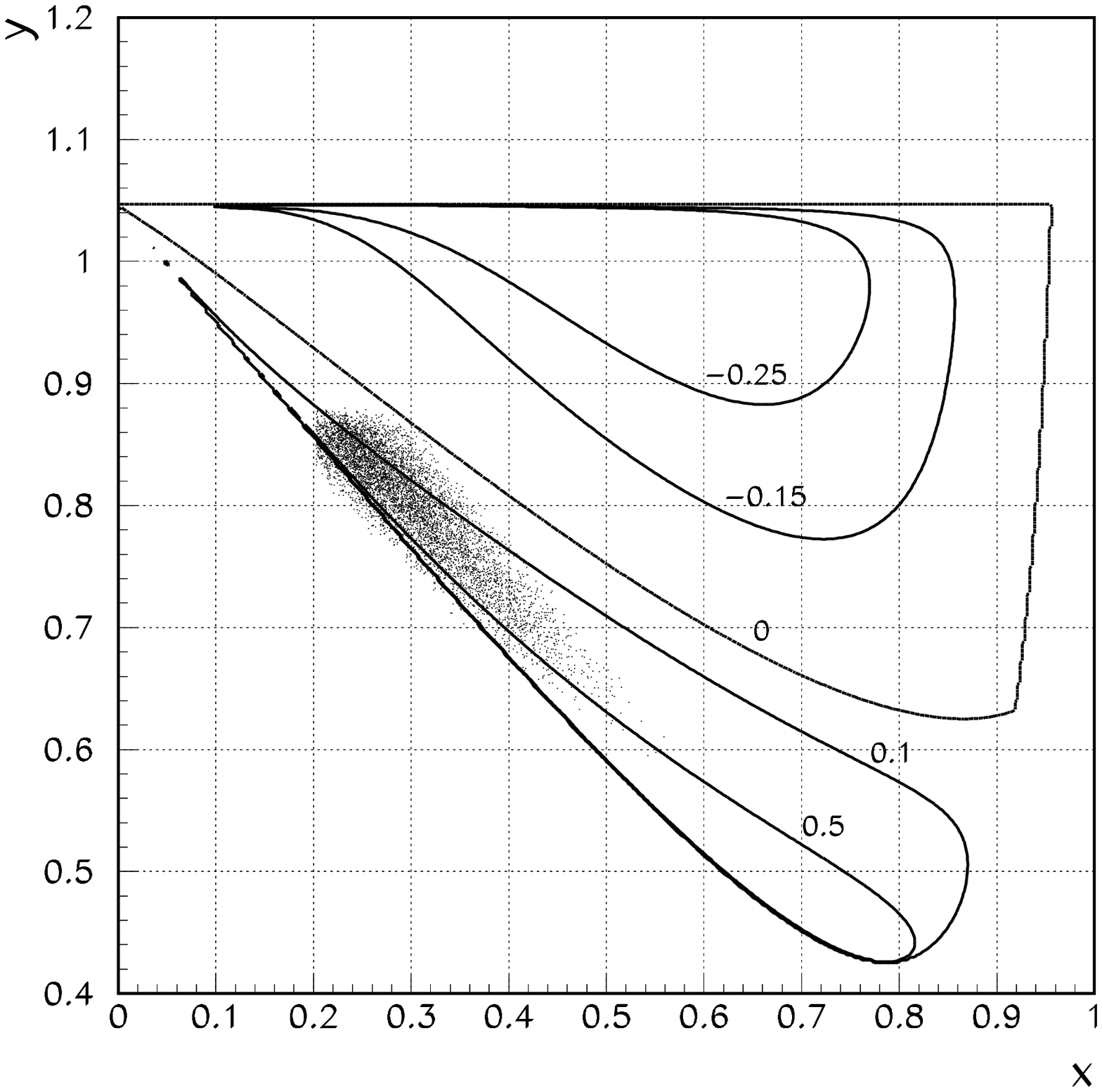}
\caption{Contour lines of the normal polarization $P_N$ 
over the $\kmng$ Dalitz plot. The dots represent the experimental 
data after tight cuts: 
$-0.7\times 10^4~{\rm MeV}^2/c^4<M^2_{miss}<1.0
\times 10^4~{\rm MeV}^2/c^4$, 
$p_{\nu}>220~{\rm MeV}/c$,
 $\Theta_{\mu\gamma}\leq 75^{\circ}$,  $E_{\gamma} > 50$ MeV;
  $x=\frac{2E_\gamma}{m_K}$ and $y=\frac{2E_\mu}{m_K}$.}
\label{pnkmng}
\end{figure}

After applying the cuts $p_{\nu}>200~{\rm MeV}/c$,
$-0.7\times 10^4~{\rm MeV}^2/c^4<M^2_{miss} < 
1.5\times 10^4~{\rm MeV}^2/c^4$ and 
$\Theta_{\mu\gamma}\leq 90^{\circ}$,  
the accepted $K_{\mu2\gamma}$  events are concentrated in 
the Dalitz plot region, as shown in Fig.~\ref{pnkmng}
where the 
 inner bremsstrahlung (IB)  
 component  and interference of the IB and structure-dependent 
with positive-helicity photon (SD$^+$) term
 dominate.   

\begin{figure}[htb]
\centering\includegraphics[width=11cm]{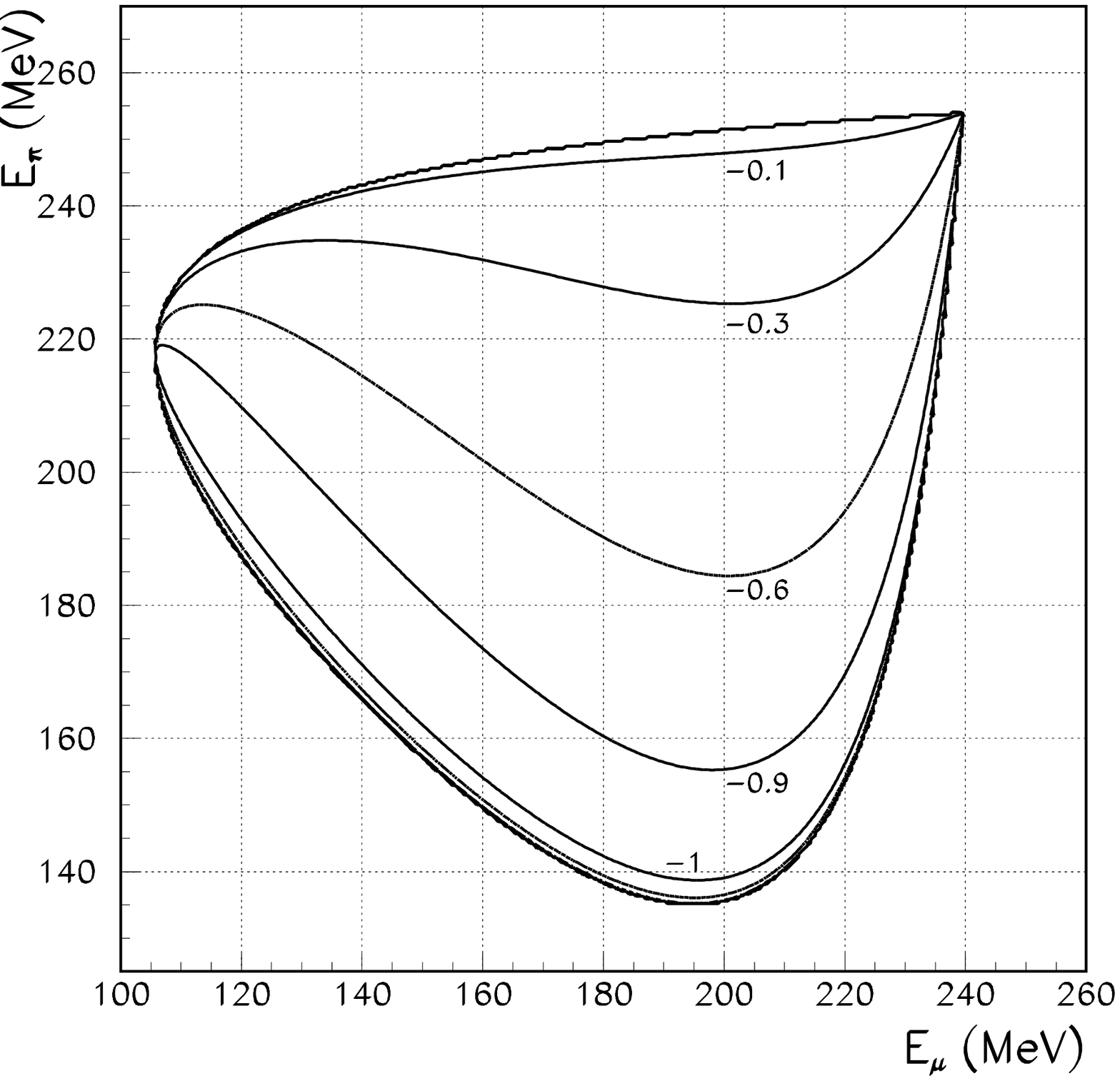}
\caption{Contour lines of $P_N$ for the $\kmnp$ Dalitz plot.}
\label{pnkmnp}
\end{figure}

A good test of the quality of detected  $\kmng$  events 
 is the value of the normal asymmetry, $A_N$. Its value is
 proportional to the T-even  muon polarization, $P_N$, which is the 
in-plane component of the muon polarization normal to the  
muon momentum 
\begin{equation}
P_N=\frac{\vec{s}_{\mu}\cdot(\vec{p}_{\mu}
\times(\vec{p}_{\gamma}\times\vec{p}_{\mu}))}
{|\vec{p}_{\mu}\times(\vec{p}_{\gamma}\times\vec{p}_{\mu})|}.
\end{equation} 
$A_N$ can be measured if 
the accepted events are separated into two classes: events 
where the photon  
moves into 
the left hemisphere with respect to the median plane of 
the given magnet 
sector and events where 
the photon  moves into the right hemisphere. The values of 
$A_N$ for these 
two classes 
should be the same but opposite in sign. 
For $\kmng$ the normal polarization has a positive sign 
in the region where 
the selected 
events are located~\cite{pt_chen}, as seen from 
Fig.~\ref{pnkmng}, and its 
average value is about 0.2,
while for  $\kmnp$, the normal muon polarization calculated using
the formulae of Ref.~\cite{pn_kmu3} has a negative sign 
over the whole 
Dalitz plot region, as shown in Fig.~\ref{pnkmnp}.
From this measurement, the value of $A_N$ for the 
one-photon $\kmnp$ events 
is $A_N({\kmnp})=(-3.87 \pm 0.06)\times 10^{-2}$ and 
for selected 
$\kmng$ events with tighter cuts $-0.7\times 10^4~{\rm MeV}^2/c^4
< M_{miss}^2 < 1.0\times 10^4~{\rm MeV}^2/c^4$, 
$p_{\nu}>220~{\rm MeV}/c$ and $\Theta_{\mu\gamma}\leq 75^{\circ}$
the normal 
asymmetry
is $A_N({\kmng})=(3.59 \pm 0.56)\times 10^{-2}$. 
These values are 
both 
non-zero and have opposite signs
consistent with the expectation.
That indicates a strong signature of 
the correct $\kmng$ selection and the suppression of 
 one-photon $\kmnp$ events. 

In Monte Carlo 
simulations  the polarization value  and corresponding asymmetry 
for  $\kmng$   was obtained using 
  the form factors from Chiral Perturbation 
Theory (ChPT) $F_V=-0.095$, 
$F_A=-0.043$~\cite{ff_chpt} and $f_K=159$ MeV.    
After applying the same cuts to the MC data we obtained 
  $A_N({\kmng})=(4.06 \pm 1.14)\times 10^{-2}$.  
These simulations were used to 
check the polarimeter sensitivity to muon polarization,  
i.e. the value of the analyzing power $\alpha$,  
and the 
geometric 
attenuation factor 
$f$. The   $\alpha$ value  obtained from $\kmng$ is in a 
good agreement with 
$\alpha = 0.289 \pm 0.015$~\cite{kmu3_new} extracted from $\kmnp$ 
MC simulation. The attenuation factor $f=0.80\pm 0.03$ was 
obtained.

From the experimental data collected during the
1996-98 beam cycles, 
a total number of $1.14\times
10^5$ $fwd + bwd$ $\kmng$ events were extracted. 
The value of the transverse asymmetry measured for the 
collected events is 
$A_T=(-0.099 \pm 0.320(stat)) \times 10^{-2}$.
The contamination of the 
beam accidental 
backgrounds in $K_{\mu2\gamma}$ 
events was estimated to be not higher than
8\%.
The constant background in the polarimeter was 
$11-12$\%.  Both these backgrounds only dilute the sensitivity 
to $P_T$, but they do not produce 
any spurious T-violating asymmetry. 
Using MC data the 
background decay
fractions 
were estimated  to be 
$\beta^{\kmnp}_{bgr}\simeq 17$\% 
and $\beta^{K_{\pi2}}_{bgr}<0.5$\%. 
Thus, the total background 
  contamination is $\beta\simeq 25$\% 
which is used in the 
denominator of Eq.~(\ref{ptat}).

Using the values of $A_T$, $\alpha$, $f$ and $\beta$ given above 
we obtained the transverse polarization
$P_T=(-0.57 \pm 1.85(stat))\times 10^{-2}$. The mean value of 
transverse polarization due to the final state interactions 
for the Dalitz plot region in which the selected $\kmng$ events 
reside is $P_T\mbox{(FSI)}=0.7\times 10 ^{-3}$ which is much less than 
the statistical error of
$P_T$. The theoretical uncertainty 
of the $P_T\mbox{(FSI)}$ value is about $15$\% and is connected 
with the uncertainties of $F_V$ and $F_A$ values. For the final 
result we subtract $P_T\mbox{(FSI)}$ from the measured $P_T$ value 
which gives $P_T=(-0.64 \pm 1.85(stat))\times 10^{-2}$ and as the 
$P_T\mbox{(FSI)}$ uncertainty is more than two orders of magnitude 
less than the statistical error we do not include that uncertainty 
in the $P_T$ error.

\begin{figure}[htb]
\centering\includegraphics[width=11cm]{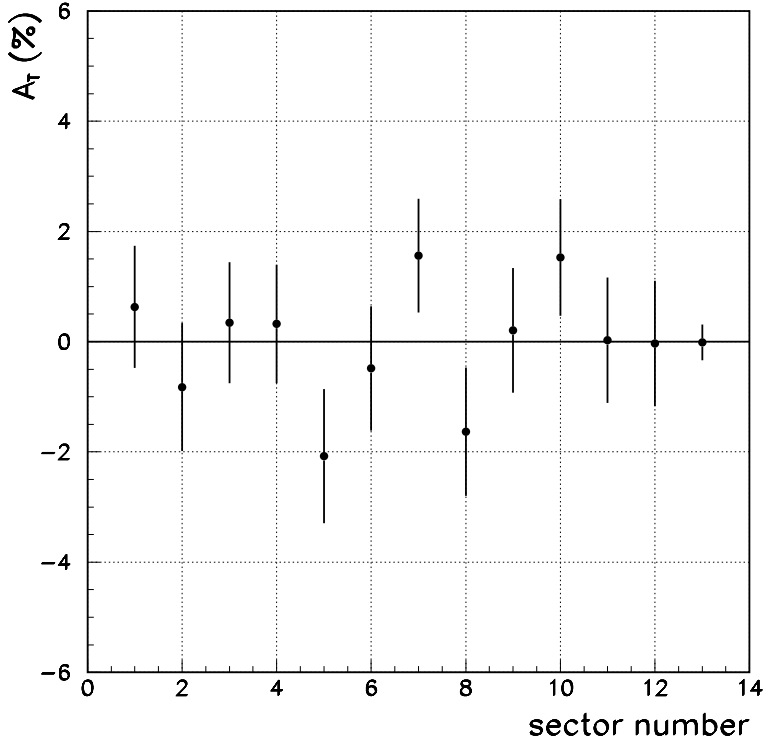} 
\caption{The dependence of the transverse asymmetry ($A_T$) 
on the sector number. 
The rightmost point represents the sum of the asymmetries 
over all 12 sectors.
The error bars show the statistical errors.}
\label{atgap}
\end{figure}

The main systematic error
contributions to $P_T$ come from 
the presence of the two large components of the  muon in-plane 
polarization, $P_L$ which is 
parallel to the muon momentum and $P_N$ ($P_T\ll P_{N,L}\leq 1$).  
However, most of their
contribution is canceled by the azimuthal symmetry of the detector 
as well as by  
the $fwd/bwd$ ratio. The largest  instrumental
systematic errors are due to the misalignment of
the polarimeter, the asymmetry of the magnetic field distribution, 
and the asymmetric
kaon 
stopping distribution. The $fwd/bwd$ ratio efficiently reduces 
these contributions. 
Additionally, a non-zero value of $P_T$ for the $\kmnp$ background 
would contribute to 
the systematic error of $P_T$ for $\kmng$, but that was measured 
in the main E246 
analysis with much higher accuracy and
is consistent with 
zero~\cite{kmu3_new}
$P_T({\kmnp})=(-1.12\pm 2.17(stat)\pm 0.90(syst))\times 10^{-3}$.
All possible sources of systematic errors  
 were
considered 
in~\cite{Abe:1999nc}
for much 
higher statistics of 1$\gamma$ $\kmnp$ events as well as for 
$K_{\mu2}$ and $K_{\pi2}$ events.
The total  systematic error of $P_T$ for $\kmnp$ events was 
found to be $0.9\times 10^{-3}$~\cite{Abe:1999nc}, while for 
$\kmng$ we estimated the systematic 
error of $P_T$ to be less than $\delta P_T^{sys}=1.0\times 10^{-3}$. 
The absence of significant systematic errors (comparable to the 
statistical error) 
is confirmed by 
measuring the dependence of the transverse asymmetry $A_T$  
on the magnet sector number as seen in Fig.~\ref{atgap}.

\section{Results}

 We have performed the first measurement  of the T-violating 
 muon polarization  in $\kmng$ decay  based on about 
 $1.14\times 10^5$ good $fwd+bwd$ $\kmng$ events collected 
 in  1996-98.  
 These events are located in the Dalitz plot region where the 
IB and  INT$^+$ components dominate.
The value obtained for $P_T$ is $(-0.64 \pm 1.85(stat) \pm 
0.10(syst))\times 10^{-2}$ including the theoretical 
background from final state interactions.
At the present level of experimental sensitivity this value 
is consistent with no T-violation 
in this decay.

\section{Acknowledgments} 

This work has been supported in Japan by a Grant-In-Aid from 
the Ministry of Education, Science, 
Sports and Culture, and by JSPS; in Russia by the Ministry of 
Science and Technology, and by the 
Russian Foundation for Basic Research; in Canada by NSERC and 
IPP, and by 
the TRIUMF infrastructure 
support provided under its NRC contribution; in the USA by 
the NSF. The authors gratefully acknowledge the support received 
from the INR and KEK staff.

\end{document}